\documentstyle[11pt,moriond,epsfig]{article}

\def\be{\begin{equation}}
\def\ee{\end{equation}}
\def\bea{\begin{eqnarray}}
\def\eea{\end{eqnarray}}
\def\munu{\mu\nu}

\begin{document}
\vspace*{4cm}
\title{PARTON SATURATION, PRODUCTION, AND EQUILIBRATION IN HIGH ENERGY 
NUCLEAR COLLISIONS}

\author{R. Venugopalan}
\address{Physics Department, BNL, Upton, NY 11973, USA}

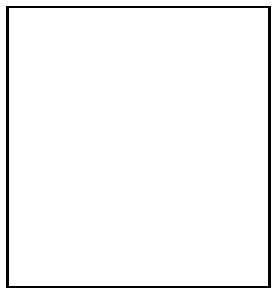
\maketitle\abstracts{Deeply inelastic scattering of electrons off
nuclei can determine whether parton distributions saturate at HERA
energies. If so, this phenomenon will also tell us a great deal about
how particles are produced, and whether they equilibrate, in high
energy nuclear collisions.}

\newpage
\section{Introduction}
The deeply inelastic scattering (DIS) experiments at HERA revealed
that structure functions grow rapidly at small x and large
$Q^2$~\cite{H1Zeus}.  At some $x_{crit}$, for a fixed $Q^2$, it is
expected that parton distributions will saturate--leading to a weaker
growth of the structure functions~\cite{GLRMQ}. Asymptotically,
at least, one expects cross--sections to not grow faster than the
logarithm squared of the energy~\cite{FroissMart}.

Recently, there have been hints in the DIS data from ZEUS that
saturation may be occuring already at HERA energies. The derivative of
the structure function $F_2$, with respect to $\log(Q^2)$, is plotted
as a function of $x$ (each bin in $x$ being averaged over the $Q^2$
acceptance for that particular $x$)~\cite{Caldwell}. The data for this
quantity, which at high $Q^2$ and relatively large $x$ may be simply
related to the gluon structure function, show the expected rise for
decreasing $x$ (and decreasing mean $\langle Q^2\rangle$), but then
flattens and decreases for very small $x$ (and small $\langle
Q^2\rangle$). An interesting feature of the data is that the turnover
region lies between $\langle Q^2 \rangle \approx$ 1--10 GeV$^2$. The 
GRV94 DGLAP fit overshoots the data in this region. However, other, more
recent parton distribution function sets do fit the data.

One must warn that the data should
be approached with some caution since the averaging procedure is
performed over a wide range of $Q^2$ for each bin. However, ZEUS has
collected more data in the low $Q^2$ region which will enable them to
perform the average over a smaller range of $Q^2$~\cite{Derrick}.  We
will assume here that atleast the qualitative features of the Caldwell
plot will remain unchanged.
 
A possible interpretation of the data--often called the ``Caldwell
Plot''--is that it demonstrates the onset of parton
saturation~\cite{MGG}. The caveat about the data aside, if
indeed we are seeing saturation at small x at HERA, this is very
relevant for nuclear collisions at LHC, and to a lesser extent for
RHIC. A sure-fire way to confirm saturation is to collide electrons with 
nuclei!~\cite{Strikman}. 

The saturation scale determines the typical intrinsic momenta
associated with quanta in the nuclear wavefunction.  These quanta go
on shell in a collision, and eventually produce a large multiplicity
of particles, mostly pions. At very high energies, the saturation
scale is the only scale in the problem, and one can estimate that the
typical momenta of the particles produced in the collision is at this
scale.  If this momenta is large enough, one can estimate in
perturbation theory whether or not they equilibrate to form a quark
gluon plasma~\cite{BlaiMuell,Muell992}.

Parton saturation, and its implications for both DIS and heavy ion
collisions, can be studied systematically in an effective field theory
(EFT) approach to QCD at small x~\cite{MV,JKMW,JKLW,JKW}. In the following
section, we discuss parton saturation in this model, both for quarks
and gluons. In section 3, we apply this model to nuclear
collisions. The very early stages of the nuclear collision, where the
non--linearities in the fields are large, can be studied in this
approach. At late times, the fields linearize. Whether
the partons, which have ``emerged'' by these times, equilibrate is an
interesting issue, of great relevance to the quark gluon plasma
community. This is discussed briefly in section 4. 

\section{Parton Saturation}

In the infinite momentum frame $P^+\rightarrow \infty$, the effective
action for the soft modes of the gluon field with longitudinal momenta
$k^{+}<<P^{+}$ (or equivalently $x\equiv k^{+}/P^{+} << 1$) can be written in
light cone gauge $A^{+}=0$ as 
\bea
S_{eff} &=& -\int d^4 x {1\over 4} G_{\munu}^{a}G^{\munu,a} +{i\over N_c}
\int d^2 x_t dx^- \rho^a(x_t,x^-)
{\rm Tr}\left(\tau^a W_{-\infty,\infty}[A^-](x^-,x_t)\right)\nonumber \\
&+& i\int d^2 x_t dx^- F[\rho^a(x_t,x^-)] \, .
\label{action}
\eea
Here $G_{\munu}^a$ is the gluon field strength tensor, $\tau^a$ are
the $SU(N_c)$ matrices in the adjoint representation, and $W$ is the
path ordered exponential in the $x^+$ direction in the adjoint
representation of $SU(N_c)$,
\be
W_{-\infty,\infty}[A^-](x^-,x_t) = P\exp\left[-ig\int dx^+
A_a^-(x^-,x_t)\tau^a\right] \, .
\ee

\newpage
The above is a general gauge invariant form~\cite{JKLW} of the
action that was proposed in Ref.~\cite{MV}.

The effective action in Eq.~\ref{action} above has a remarkable saddle point
solution~\cite{MV,JKMW,Kovchegov}. It is equivalent to solving the Yang--Mills
equations
in the presence of the source $J^{\mu,a} = \delta^{\mu +}\rho^a(x_t,x^-)$.
One finds a solution where $A^{\pm}=0$, and
$A^i = {-1\over {ig}}\,V\partial^i V^\dagger$  
(for $i = 1,2$)
is a pure gauge field which satisfies the equation
$D_i {d A^i\over dy} = g \rho (y,x_\perp)$. 
Here $D_i$ is the covariant derivative $\partial_i + V\partial_i
V^\dagger$, $y=y_0 + \log(x^-/x_0^-)$ is the space--time rapidity, and
$y_0$ and $x_0$ are the space-time rapidity and the Lorentz contracted
width, respectively, of the hard partons in the fragmentation
region. At small x, the space--time and momentum space notions of
rapidity are used interchangeably~\cite{RajLar}.  The momentum space
rapidity is defined to be $y = {\tilde{y}}_0 - ln(1/x)$ where $x$ is
Bjorken $x$ and ${\tilde{y}_0}$ is the rapidity in the fragmentation
region.  The solution of the above equation is~\cite{JKMW}
\be
A_\rho^i(x_t) = {1\over ig}\left(Pe^{ig\int_y^{y_0}
dy^\prime
{1\over {\nabla_{\perp}^2}}\rho(y^\prime,x_t)}\right)
\nabla^i\left(Pe^{ig\int_y^{y_0} dy^\prime {1\over
{\nabla_{\perp}^2}}\rho(y^\prime,x_t)}\right)^\dagger \, .
\label{puresoln}
\ee

To compute the classical nuclear gluon distribution function, one
needs in general to average over the product of the classical fields
in Eq.~\ref{puresoln} at two space--time points with the weight
$F[\rho]$~\cite{MV}.  For a large nucleus, one may approximate
$F[\rho]\longrightarrow \int d^2 x_t dy {1\over \mu^2(y)}{\rm Tr}
\left(\rho^2\right)$, 
where $\mu^2$ is the color charge squared per
unit area per unit rapidity. The classical gluon distribution for this
Gaussian source is
\be
{dN\over d^2 x_t} = {1\over 2\pi\alpha_S} {C_F\over 
x_t^2} \left( 1-\exp\left( -{\alpha_S\pi^2\over 2\sigma C_F} x_t^2 xG\left(x,{1
\over x_t^2}\right)\right) \right)\, ,
\label{glue}
\ee 
where $C_F$ is the Casimir in the fundamental representation and 
$\sigma$ is the nuclear cross-section~\footnote
{ Above, we have re-written the expression for the gluon distribution
in Ref.~\cite{JKMW}, using the leading log gluon distribution to
replace $\mu^2$ and $\log(x_t\Lambda_{QCD})$ with the gluon distribution 
$xG(x,{1\over x_t^2})$ at the scale $1/x_t^2$.}. 

For large $x_t$ (but smaller that $1/\Lambda_{QCD}$, the distribution
falls like a power law $1/x_t^2$--and has a $1/\alpha_S$ dependence! 
For very small $x_t$, the behavior
is the perturbative distribution $\log(x_t \Lambda_{QCD})$. The scale
which determines the cross--over from a logarithmic to a power law
distribution is, to follow the notation of Mueller~\cite{Muell991}, the
saturation scale $Q_s$. Setting $x_t=1/Q_s$ and the argument of the 
exponential above to unity, one obtains the relation, 
\be
Q_s^2 = {\alpha_S \pi^2\over 2\sigma}{1\over C_F} xG(x,Q_s^2) \, ,
\label{gluesat}
\ee
which, for a particular $x$, can be solved self--consistently to
determine $Q_s$. The value of $Q_s$ is approximately 1 GeV for RHIC energies 
and 2--3 GeV for LHC energies. To compare with the estimate of 
Gyulassy--McLerran~\cite{GyuMc}, set $Q_s=C\alpha_S\sqrt{\chi}$, 
in the notation of Ref.~\cite{MV,JKMW} where 
$\chi = \int_y^{{\tilde y}_0} \mu^2(y)$ and $C\sim 5$ for $Q_s\sim 1$ GeV.

The model calculation above shows that a) the gluon 
distribution saturates at some scale $Q_s >> \Lambda_{QCD}$--the
momentum distribution of partons grows only logarithmically for $k_t<
Q_s$, and b) this saturation is seen already at the classical level. 
The gluon distribution has a tail that goes as $1/k_t^2$, so one expects 
the typical intrinsic momentum of the gluons to be peaked at $k_t\sim Q_s$.

A similar behavior is seen for the structure function $F_2$. The quark
distribution in the classical gluon field-and $F_2$, can be computed in
the same approach~\cite{MV98,KKMV}, and one obtains
\bea
F_2(x,Q^2) &=& {Q^2\sigma N_c\over {2\pi^3}} 
\sum_f e_f^2 \int_0^1 dz \int_0^{1\over \Lambda_{QCD}} dx_t x_t 
\left(1-\exp\left(-{\alpha_S\pi^2\over 2\sigma C_A}x_t^2 xG(x,{1\over x_t^2})
\right)\right) \, \nonumber \\
&\times& \left[K_0^2 (x_t A_f)\left(4z^2(1-z)^2 Q^2 +M_f^2\right)+ 
K_1^2(x_t A_f)A_f^2 \left(z^2+(1-z)^2\right)\right] \, ,
\label{Glauber}
\eea
where $e_f^2$ is the electric charge squared of a quark of flavor $f$,

\newpage
$A_f^2 = Q^2 z(1-z) + M_f^2$, $K_{0,1}$ are the modified Bessel
functions, and $C_A$ is the Casimir in the adjoint representation. 
The above equation is the well known Glauber expression~\cite{NikZak}
usually derived in the rest frame of the nucleus. It is heartening
that, with the assumption of Gaussian color charges~\cite{KKMV}, the
formalism of Ref.~\cite{MV98} for structure functions in the infinite 
momentum frame (which,
 in principle also accomodates non--linear evolution for structure
functions) reproduces it.  For large $Q^2$, it reduces to the standard
DGLAP expression, while at small $Q^2$ it goes to zero as $Q^2
log(Q^2)$. One then recovers, qualitatively, the shape of the Caldwell
plot for $dF_2/d\log(Q^2)$.

One obtains from the above equation, in a manner analogous to
Eq.~\ref{glue}, the quark saturation scale $Q_s^q$ by replacing
$C_F\longrightarrow C_A$ in Eq.~\ref{gluesat}.  The relative size of
the two saturation scales, glue to quark, is therefore determined
simply by the ratio of the two Casimirs, $C_A/C_F$.

What about quantum corrections to the above quark and gluon
distributions?  At the one loop level, one gets $\log(1/x)$
corrections to the Weizsacker--Williams
distribution~\cite{Muell90,JAMV,Muell991}. However, Mueller has argued
recently that beyond the one loop level, the distribution has the same
form as the as the above classical gluon distribution. What does
change due to small $x$ evolution is the $x$ dependence of the
saturation scale~\cite{Muell991}.  It will be very interesting to see
if detailed studies of parton evolution in the non--linear region
confirm this result~\cite{Yuri,JKW}.

\section{Parton Production}

In the previous section, we discussed the parton distribution in a
single nucleus in an effective field theory approach to QCD at small
x.  Kovner, McLerran, and Weigert applied this approach to study
nuclear collisions at very high energies~\cite{KMW}. One now has two
sources of color charge $\rho(x^\pm)$, one on each light cone,
described by the currents $J^{\pm,a}=\rho_{\pm}^a(x_t)\delta(x^{\mp})$
corresponding to $P^{\pm}\rightarrow \infty$. The classical field
describing the small x modes in the EFT is then obtained by solving
the Yang--Mills equations in the presence of these sources. It can be
written as ${\tilde A}^\mu = \theta(x^+)\theta(x^-)A^\mu +
\theta^(x^-)\theta(-x^+)A_1^\mu +\theta(-x^-)\theta(x^+)A_2^\mu$.
Here $A_{1,2}^\mu$ have $A^\pm=0$ and the transverse gauge fields
$A_{1,2}^i$ are, as discussed in the previous section, pure gauges.
 
A convenient co--ordinate system to study nuclear collisions is $x^\mu
= (\tau,\eta,{\vec x}_\perp)$, where $\tau = \sqrt{2x^+ x^-}$ and
$\eta = {1\over 2} \log({x^+\over x^-})$. 
Matching the Yang--Mills
equations in the different light cone regions at $\tau=0$, one can
eliminate singular terms by requiring $A^i = A_1^i+A_2^i$ and $A^\pm =
\pm x^\pm {ig\over 2}[A_1^i, A_2^i]$ at $\tau =0$ ($i=1,2$ are the two
transverse co--ordinates). The sum of two pure gauges in QCD is {\it
not} a pure gauge. The coherence of the Weizs\"acker--Williams fields
is broken, and particles are produced--driven by the above commutator
term.

Kovner, McLerran, and Weigert solved the Yang--Mills equations
perturbatively to $O(\rho^2)$. Diagrammatically, this can be
represented as $qq\longrightarrow qqg$. In Schwinger gauge,
$A^\tau=0$, and for $k_t>>\alpha_S\mu$, this is the dominant
contribution to gluon production at small x. The result has been
computed in different ways by several
authors~\cite{KMW,KRetal} and it agrees with the
pQCD gluon bremsstrahlung expression of Gunion and Bertsch at small
x~\cite{GunBert}.

To the order $O(\rho^2)$, the saturation effects discussed in the
previous section are not visible. The result agrees with the pQCD
mini--jet result at small x. They become less reliable below the scale
$k_t\sim Q_s$, where all orders in $\rho$ become important. Given the
distribution in Eq.~\ref{glue}, suggesting that most of the partons in
the small x component of the nuclear wavefunction have intrinsic
momenta $k_t\sim Q_s$, these higher order effects are important for
nuclear collisions. They self-consistently regulate the divergence at
small $k_t$, leading to infrared safe distributions of produced
particles. Saturation effects may only be marginal at RHIC, but will
be absolutely essential at LHC-where most of the produced particles
will be semi--hard.

Recently, Krasnitz and I computed gluon production to all orders in
$\rho$ numerically~\cite{AlexRaj1,AlexRaj2}. Gauge fields in the
forward light cone were obtained by solving Hamilton's equations, for
each $\rho$ configuration, on a lattice. Observables were computed by
averaging over these configurations with a Gaussian weight. Assuming
boost invariance, the lattice Hamiltonian is the Kogut--Susskind
Hamiltonian in 2+1--dimensions, coupled to an adjoint scalar
field. 

\newpage
The lattice initial conditions, in analogy with the continuum
case, were obtained by matching the lattice equations of motion in the
different light cone regions at $\tau=0$. Our study was performed on 
two dimensional transverse lattices ranging from 10$\times$10 to 
160$\times$160. For simplicity, we studied only the SU(2) case.

For large transverse momenta $k_t\gg \alpha_S\mu$, the results of our
simulation agree very well with lattice perturbation theory. This is
the case for both the momentum and time dependence of the fields. The
lattice coupling is $g^2\mu L$. For $g^2\mu L\gg 1$, one observes
large non--perturbative corrections at smaller momenta. At very early
times, the field strengths are very large--and a parton model
description is likely not valid. Since $\alpha_S\mu$ (or $Q_s$) is the
only scale in the problem, one expects fields to linearize on the time
scale $\tau_{l}\sim 1/\alpha_S\mu$. This is seen when we compute $\epsilon 
\tau$ on the lattice. It rises from zero and goes to a constant at $\approx 
1/\alpha_S\mu$. At late times $\tau\sim t$, this  quantity is the energy
per unit transverse area per unit rapidity of produced particles. 
It scales as $\mu^3$.

To summarize the above, we now have the technology to systematically
compute a range of gauge invariant observables in high energy heavy
ion collisions. An interesting quantity to compute would be the number
of particles produced per unit rapidity. Several years ago, Blaizot
and Mueller~\cite{BlaiMuell}--assuming saturation, computed on fairly 
general grounds the number of produced gluons per unit rapidity to be 
\be
{dN\over dy} = c\,\, 2A xG(x,Q_s^2) \, ,
\ee
with $c=1$. As Mueller has pointed out recently~\cite{Muell992}, this
coefficient cannot be computed precisely without knowing the details
of the collision. It is now feasible for us to confirm this formula
and compute the value of the coefficient $c$~\footnote{The author thanks 
L. McLerran for discussions on this point}. We hope to report on
it in the near future.

If saturation does occur at $Q_s\sim 1$ GeV and $x\sim$ 10$^{-4}$, as the 
HERA data suggest, then depending on the x dependence of the gluon density, 
saturation may occur already for large nuclei at $x\sim$ 10$^{-2}$--which 
corresponds to RHIC energies. Modulo what $c$ above is, the multiplicity of 
produced particles at RHIC may tell us something about the saturated gluon 
density at HERA or vice versa. At LHC, the gluon densities would correspond 
to lower values of x for a fixed $Q^2$ than at HERA (hence the urgent 
need for an eA collider!).

\section{Parton Equilibration}

At very early times after the nuclear collision, non--linearities in
the Yang--Mills equations are extremely important, and the concept of
partons is not very meaningful. The non--linearities
dissipate-producing on shell partons in a time scale $\tau\propto
1/\alpha_S\mu$. Since this time scale is much less than $\tau = R$, it
is likely the produced partons will further interact, and perhaps
equilibrate. (It is only at this stage that it is meaningful to think
of a parton cascade.) The further interaction of produced partons is
beyond the scope of the simulations of Krasnitz and myself. However,
the ``final'' distribution of partons in our model can, in principle,
be used as the initial conditions for a parton cascade. We should also
point the reader to the Mueller's recent qualitative study of small
angle scattering and the onset of equilibration in heavy ion
collisions~\cite{Muell992}. From the theoretical viewpoint, the
problem of parton equilibration is an extremely difficult one--especially 
since it is not clear that the problem can be formulated in a gauge invariant 
way.

\section*{Acknowledgments}
I would like to thank Prof. Tran Thanh Van and the other organizers for a 
very well organized and stimulating conference. I would also like to 
thank the US Moriond committee for an NSF travel award. This work has been 
supported by DOE nuclear theory at BNL.

\end{document}